\documentclass[a4paper,12pt]{article}
\usepackage{amsmath,amsthm,amssymb,bbm}

\usepackage[english]{babel}

\title{\bf Bipartite entanglement in systems of identical particles:
the partial transposition criterion}

\author{F. Benatti$^{a,b}$, 
R. Floreanini$^{b}$, U. Marzolino$^{c}$\\
\\
\small ${}^a$Dipartimento di Fisica, Universit\`a di Trieste, 
34151 Trieste, Italy\\
\small ${}^b$Istituto Nazionale di Fisica Nucleare, Sezione di Trieste,
34151 Trieste, Italy\\
\small ${}^c$Universit\`a di Salerno, 
84084 Fisciano, Italy}

\date{\null}

\begin{document}

\maketitle

\begin{abstract}
\noindent
We study bipartite entanglement in systems of $N$ identical bosons distributed
in $M$ different modes. For such systems, a definition of separability not related 
to any {\it a priori} Hilbert space tensor product structure is needed and can be given
in terms of commuting subalgebras of observables. Using this generalized notion
of separability, we classify the states for which partial transposition
turns out to be a necessary and sufficient condition for entanglement detection.
\end{abstract}

\section{Introduction}

The characterization and quantification of non-classical correlations have become one
of the most important tasks in quantum physics.%
\footnote{See \cite{Horodecki} and references therein.}
Quantum correlations, besides allowing
the implementation of classically unavailable protocols in information theory, are emerging
as a fundamental tool in explaining the behaviour of many quantum systems, specifically
those containing a large number of elementary constituents for which conventional
approaches fail to provide a satisfactory description \cite{Lewenstein,Bloch}.

Most of the studies devoted to the understanding and measure of the entanglement
present in many-body systems have been focused on spin systems and the like,
{\it i.e.} on systems where the microscopic constituents can be singly addressed.
In this case, entanglement theory is well developed and various tests
able to recognize and quantify non-classical correlations in such systems states
have been identified and widely used in applications \cite{Horodecki,Amico}.

The situation is less definite and clear when dealing with many-body systems made of
identical particles, where the microscopic constituents can not be singly addressed and
their properties measured: only collective, global operators are in fact admissible
observables~\cite{Feynman,Sakurai}. Such systems are becoming more and more relevant in
many applications, thanks to the recent advances in quantum optics,
confined ultracold gases and superconducting systems, and more in general in the physics
of mesoscopic systems and quantum devices.

For systems in which the microscopic constituents are indistinguishable, the usual
definitions of separability and of entanglement are inapplicable since the natural
particle tensor product structure on which these notions are based is no longer
available. More specifically, in a system made of $N$ distinguishable constituents
({\it e.g.} qubits, spins, quantum dots...), 
a generic state $\omega$ ({\it i.e.} a positive
linear functional on the Banach space of the observables) is said to be entangled
if it can not be written as a convex combinations of product states, namely
\begin{equation}
\omega=\sum_k p_k\, \omega_k^{(1)}\otimes\omega_k^{(2)}\otimes\ldots\otimes\omega_k^{(N)}\ ,
\qquad p_k\geq 0\ ,\quad \sum_k p_k=1\ ,
\label{1}
\end{equation}
where $\omega^{(i)}$ represents a state for the $i$-th constituent, with the
associated ``single-particle'' Hilbert space ${\cal H}^{(i)}$.
This definition stems from the natural splitting of the total Hilbert space $\cal H$
of the system in the tensor product of the single constituent Hilbert spaces:
\begin{equation}
{\cal H}={\cal H}^{(1)}\otimes{\cal H}^{(2)}\otimes\ldots\otimes{\cal H}^{(N)}\ .
\label{2}
\end{equation}
On the other hand, systems made of identical constituents can not be described by states
as in (\ref{1}) since the Hilbert space $\cal H$ must now contain only elements that are
even (for bosons) or odd (for fermions) under any permutation of the elementary constituents.
This fact points to the need of a generalized notion of separability and entanglement,
not based on the tensor decomposition in (\ref{2}), or more in general on
the ``particle'' aspect of first quantization. The emphasis should shift from the
set of system states to the corresponding algebra of observables and to the associated
correlation functions, {\it i.e.} to a view closer to the second quantization 
description of many-body systems \cite{Zanardi1}-\cite{Argentieri}.

Following this framework, a new definition of separability has been proposed in \cite{Benatti}
that applies equally well to all situations and reduces to the one given in (\ref{1})
for systems of distinguishable particles. Although the notion of entanglement
in many-body systems has been widely discussed in the recent literature
\cite{Schliemann}-\cite{Buchleitner},
only limited part of those results are really applicable to the case of identical particles.
We stress that the definition introduced in \cite{Benatti} is instead very general,
suitable for all physical situations; in particular, it has been applied to the study of trapped
ultracold bosons, leading to novel experimental testable predictions 
in quantum metrology \cite{Benatti1,Argentieri}.

Having a general, consistent notion of entanglement is however not enough
for a practical use in applications; as in the case of the standard definition based
on the tensor decomposition (\ref{2}), one needs to identify criteria able to detect non-classical
correlations through easily implementable tests \cite{Horodecki,Bengtsson}.

In the case of bipartite entanglement in systems of distinguishable qubits, the operation of 
partial transposition has been identified as one of such criteria
\cite{Peres,Horodecki2}: the negativity of the partially 
transposed state is a sure sign of the presence of quantum correlations. Although in general
a sufficient condition for bipartite entanglement, 
the partial transposition criteria becomes also necessary in lower dimensions,
either for a $2\times 2\,$-- or $2\times 3\,$-- dimensional Hilbert space \cite{Peres,Horodecki2}, 
or, in the case of continuous variable models, for bipartite Gaussian states in which 
one part contains just one mode \cite{Simon,Werner}.

In the following, we shall discuss to what extent these results can be extended to bosonic systems
made of a fixed number of elementary constituents.
Using the generalized definition of separability and the language of second quantization,
we shall classify the states for which the partial transposition operation results
a necessary and sufficient criterion for detecting bipartite entanglement. As a byproduct of
these results, a complete characterization of the states that remain positive
under the operation of partial transposition will also be given, a fundamental step
towards the full classification of bound entangled states in such bosonic systems.

\section{Entanglement and identical particles}

As remarked in the Introduction, for a generic many-body system made of distinguishable particles
the usual notion of separability, 
as expressed by (\ref{1}), makes use of the particle
tensor product structure (\ref{2}) of the system Hilbert space,
a clear heritage of the intrinsic addressability of the single elementary constituents:
it reflects the natural multi-partition of the system into its elementary parts.

On the other hand, states of the form (\ref{1}) are not allowed
states for systems made of identical particles; in fact, 
assigning a state $\omega^{(1)}$ to the first particle, $\omega^{(2)}$
to the second and so on, would imply the possibility of distinguishing
them through their states. Instead, the rules of quantum mechanics
assign to a system of identical particles a total Hilbert space $\cal H$
made of elements that are totally symmetric (for bosons) or antisymmetric
(in the case of fermions) under any permutation of the particles; similarly,
also the system observables must correspondingly be either even
or odd under the same class of permutations \cite{Feynman,Sakurai}.

As a consequence, the notion of separability based on the decomposition (\ref{1}) is no longer
applicable in the case of bosonic (or fermionic) many-body systems:
a suitable generalization encompassing it is needed.
The basic idea is to give emphasis to the set of the system observables
instead of that of its physical states; these two ``points of view'' are
``dual'' to each other: the connection between the two is given by the expectation value map
that allows to express the average of an observable ${\cal O}$ as the value taken
by the system state $\omega$ on it, $\langle{\cal O}\rangle=\omega({\cal O})$.%
\footnote{A standard representation of this expectation value map
is given by the trace operation over density matrices.}
Recalling \cite{Benatti}, we shall first give an abstract definition of this generalized notion of separability
and then apply it to the specific case of multi-mode bosonic systems.

Let us consider a many-body system described by the Hilbert space ${\cal H}$ and denote by
${\cal B}({\cal H})$ the algebra of all bounded operators on it; the observables 
of the system are part of this algebra. We shall introduce the notion of 
(bipartite) separability by considering couple of commuting subalgebras of ${\cal B}({\cal H})$
instead of focusing on partitions of its Hilbert space. 
We then introduce the following preliminary definitions:

\medskip
\noindent
{\bf Definition 1.} {\sl An {\rm algebraic bipartition} of the algebra ${\cal B}({\cal H})$ is any pair
$({\cal A}_1, {\cal A}_2)$ of commuting subalgebras of ${\cal B}({\cal H})$,
${\cal A}_1, {\cal A}_2\subset {\cal B}({\cal H})$; this means that any
element of the subalgebra ${\cal A}_1$ commutes with each element of ${\cal A}_2$, in short: 
$[{\cal A}_1, {\cal A}_2]=\,0$.}
\medskip

\noindent
The two subalgebras ${\cal A}_1$ and ${\cal A}_2$ need not reproduce the whole
algebra ${\cal B}({\cal H})$, {\it i.e.} in general 
\hbox{${\cal A}_1 \cup {\cal A}_2\subset {\cal B}({\cal H})$}.
In this respect, the term ``bipartition'' used above
should be interpreted in loose terms; nevertheless, in the cases discussed below, the considered mode
partitions actually generate the whole algebra ${\cal B}({\cal H})$.
The notion of algebraic bipartition allows defining the system local observables:

\medskip
\noindent
{\bf Definition 2.} {\sl An element (operator) of ${\cal B}({\cal H})$ is said to be {\sl local} with respect to
a given bipartition $({\cal A}_1, {\cal A}_2)$ if it is the product $A_1 A_2$ of an element 
$A_1$ of ${\cal A}_1$ and another $A_2$ in ${\cal A}_2$.}

\medskip
\noindent
We can now introduce the generalized notion of separability and entanglement:

\medskip
\noindent
{\bf Definition 3.} {\sl A state $\omega$ on the algebra ${\cal B}({\cal H})$ will be called {\sl separable} with
respect to the bipartition $({\cal A}_1, {\cal A}_2)$ if the expectation $\omega(A_1 A_2)$ 
of any local operator $A_1 A_2$ can be decomposed into a linear convex combination of
products of expectations:
\begin{equation}
\omega(A_1 A_2)=\sum_k\lambda_k\, \omega_k^{(1)}(A_1)\, \omega_k^{(2)}(A_2)\ ,\qquad
\lambda_k\geq0\ ,\quad \sum_k\lambda_k=1\ ,
\label{3}
\end{equation}
where $\omega_k^{(1)}$ and $\omega_k^{(2)}$ are states on ${\cal B}({\cal H})$;
otherwise the state $\omega$ is said to be {\sl entangled} with respect the bipartition
$({\cal A}_1, {\cal A}_2)$.}
\medskip

This generalized definition of separability 
results meaningful in all situations and can be easily extended to the case
of more than two partitions, by an appropriate, straightforward generalization
of {\sl Definition 1} and {\sl Definition 2}; specifically, in the case of an $n$-partition, Eq.(\ref{3})
would extend to:
\begin{equation}
\omega(A_1 A_2\cdots A_n)=\sum_k\lambda_k\, \omega_k^{(1)}(A_1)\, \omega_k^{(2)}(A_2)\cdots
\omega_k^{(n)}(A_n)\,\ ,\quad
\lambda_k\geq0\ ,\quad \sum_k\lambda_k=1\ .
\label{4}
\end{equation}
Further, when dealing with systems of distinguishable particles, {\sl Definition 3} gives
the standard notion of separability ({\it cf.} Eq.(\ref{1})). In this case, the partition of
the system into its elementary constituents induces a natural tensor
product decomposition not only of the Hilbert space $\cal H$ as given in (\ref{2}), but also of the
algebra ${\cal B}({\cal H})$ of its operators; a direct application of the condition (\ref{4}) to
this natural tensor product multipartition immediately yields the decomposition (\ref{1}).

In this respect, it should be noticed that when dealing with systems of identical particles, 
there is no {\it a priori} given, natural partition,
so that questions about entanglement and separability, non-locality and locality 
are meaningful only with reference to a specific choice of commuting algebraic sets of observables; 
this general observation, often overlooked in the literature, is at the basis of the definitions
given in (\ref{3}) and (\ref{4}).

A special situations is represented by pure states. In fact, when dealing with pure states 
instead of general statistical mixtures and bipartitions that involve the whole algebra
${\cal B}({\cal H})$, the separability condition in (\ref{3}) (and similarly for (\ref{4}))
simplify, becoming:
\begin{equation}
\omega(A_1 A_2)=\omega^{(1)}(A_1)\, \omega^{(2)}(A_2)\ ;
\label{5}
\end{equation}
in other terms, separable, pure states are just product states. 
Indeed, recall that a state $\omega$ is said to be pure if it can not be expressed 
as a convex combination of other states; therefore, the sum in (\ref{3}) must contain
just one term, otherwise it would be in contradiction with the hypothesis of
$\omega$ being pure. Further, by taking either $A_1$ or $A_2$ to be the identity operator,
from the expression in
(\ref{5}) one sees that the factor states $\omega^{(1)}$ and $\omega^{(2)}$ coincide with
the restrictions of $\omega$ to the first and second partition, respectively,
so that (\ref{5}) reduces to:
\begin{equation}
\omega(A_1 A_2)=\omega(A_1)\, \omega(A_2)\ .
\label{5-1}
\end{equation}
This observation will be useful in the following.

\section{Multimode boson systems}

Henceforth, we shall focus on bosonic many-body systems, whose elementary constituents
can be found in $M$ different states or modes. This is a very general framework, useful for the
description of physical systems in quantum optics, in atom and condensed matter physics.
Among the many models fitting this paradigm, those consisting of ultracold gases of bosonic 
atoms confined in multi-site optical lattices is of special relevance. 
They turn out to be a unique laboratory, both theoretically and experimentally, 
for the study of quantum effects in many-body physics, {\it e.g.} in quantum phase transition
and matter interference phenomena, and also for applications in quantum information 
({\it e.g.} see \cite{Stringari}-\cite{Yukalov}, and references therein).

In dealing with (bosonic) many-body systems, it is most appropriate to adopt
the second quantized formalism. Let us thus introduce creation $a^\dagger_i$ and annihilation 
operators $a_i$, $i=1, 2,\ldots,M$, for the $M$ different modes that the bosons can occupy;
they obey the standard canonical
commutation relations, $[a_i,\,a^\dagger_j]=\delta_{ij}$, and 
generate out of the vacuum state $|0\rangle$ single-particle orthonormal basis states 
$|i\rangle\equiv a^\dagger_i |0\rangle$. The total Hilbert space $\cal H$ of the system
is then spanned by the many-body Fock states,
\begin{equation}
|n_1, n_2,\ldots,n_M\rangle= {1\over \sqrt{n_1!\, n_2!\cdots n_M!}}
(a_1^\dagger)^{n_1}\, (a_2^\dagger)^{n_2}\, \cdots\, (a_M^\dagger)^{n_M}\,|0\rangle\ ,
\label{6}
\end{equation}
where the integers $n_1, n_2, \ldots, n_M$ are the occupation numbers of the different modes.
In this language, symmetrization of the elements of $\cal H$, as required by identity of the particles
filling the $M$ modes, is automatically guaranteed by the commutativity of the creation operators.

Furthermore, notice that the set of polynomials in all creation and annihilation operators,
$\{a^\dagger_i,\, a_i\}$, $i=1,2,\ldots, M$,
form an algebra that, together with its norm-closure, coincides with the algebra
${\cal B}({\cal H})$ of bounded operators of the $M$-mode bosonic system.%
\footnote{Strictly speaking, the polynomials themselves are not bounded operators
(nor are the single creation and annihilation operators); the standard way to 
properly define them is through the introduction of the Weyl operators and the corresponding
algebra of bounded operators, from which polynomial operators can be defined by differentiation
\cite{Thirring,Strocchi}.}

A bipartition of this algebra is determined by an integer $m$, $0\leq m \leq M$,
splitting the collection of creation and annihilation operators into two disjoint sets
$\{a_i^\dagger,\, a_i\, | i=1,2\ldots,m\}$ and 
$\{a_\alpha^\dagger,\, a_\alpha,\, |\, \alpha=m+1,m+2,\ldots,M\}$; 
all polynomials in the first set (together with their norm-closures)
form a subalgebra ${\cal A}_1$, while the remaining set analogously generates
a subalgebra ${\cal A}_2$. Since operators pertaining to different modes
commute, one sees that $[{\cal A}_1,\, {\cal A}_2]=\,0$ and thus
the pair $({\cal A}_1,\, {\cal A}_2)$ indeed forms an algebraic bipartition
of the whole algebra ${\cal B}({\cal H})$.%
\footnote{
There is no loss of generality in assuming the modes forming the subalgebras ${\cal A}_1$
(and ${\cal A}_2$) to be contiguous; if in the chosen bipartition this is not the case, 
thanks to the commutativity of operators pertaining to different modes,
one can always relabel the modes in such a way to achieve such convenient ordering.
}
For sake of clarity, we shall use latin indices to label operators in the
set ${\cal A}_1$, while adopting greek indices for those in ${\cal A}_2$
and similarly for the mode occupation numbers (see below).

As explained in the previous Section, from an algebraic bipartition $({\cal A}_1,\, {\cal A}_2)$
one inherits the notion of locality for the elements of ${\cal B}({\cal H})$ (see {\sl Definition 2})
and consequently that of separability and entanglement (see {\sl Definition 3}).

The case in which the two commuting algebras ${\cal A}_1$ and ${\cal A}_2$ are generated
only by a subset $M'<M$ of modes can be similarly treated. Indeed, 
all operators in ${\cal B}({\cal H})$ pertaining to the modes not involved in the bipartition
commute with any element of the two subalgebras ${\cal A}_1$ and ${\cal A}_2$,
and therefore effectively act as ``spectators''. As a consequence, all the considerations
and results discussed below holds also in this situation, provided one replaces
the total number of modes $M$ with $M'$, the actual number of modes used 
in the chosen bipartition.

Given our bosonic, $M$-mode many-body system, we shall first consider pure states 
and give a complete characterization of the separable ones, through the following:

\medskip
{\bf Proposition 1.} {\sl A pure state $|\psi\rangle$ is separable with respect of the
above $({\cal A}_1,\, {\cal A}_2)$-bipartition ({\it i.e.} it is 
$({\cal A}_1,\, {\cal A}_2)$-separable) if and only if it is generated 
out of the vacuum state by a $({\cal A}_1,\, {\cal A}_2)$-local operator, in short
\begin{equation}
|\psi\rangle={\cal P}(a^\dagger_1, \ldots, a^\dagger_m)\cdot 
{\cal Q}(a^\dagger_{m+1},\ldots ,a^\dagger_M)\ |0\rangle\ ,
\label{7}
\end{equation}
where ${\cal P}$, ${\cal Q}$ are polynomials in the creation operators
relative to the first $m$ modes, the last $M-m$ modes, respectively.
By Definition 3, a pure state $|\psi\rangle$ that can not be cast in the form (\ref{7})
is therefore $({\cal A}_1,\, {\cal A}_2)$-entangled.}
\medskip 

\noindent
{\sl Proof:} First of all recall that in the present situation the condition of
separability reduces to the simpler expression (\ref{5}); clearly, the state
in (\ref{7}) satisfies it by taking for the expectation value of a generic
operator $A\in {\cal B}({\cal H})$
the usual state-average: $\omega(A)\equiv\langle\psi| A|\psi\rangle$.
In order to prove the converse, {\it i.e.}
that from the separability condition (\ref{5}) the expression (\ref{7}) follows,
let us decompose $|\psi\rangle$ in the Fock basis given in (\ref{6}):
\begin{equation}
|\psi\rangle=\sum_{\{n\}} C_{\{n\}}\ | n_1, n_2,\ldots, n_M\rangle\ ,\qquad
\sum_{\{n\}}  |C_{\{n\}}|^2 =1\ ,
\label{8}
\end{equation}
or equivalently, taking into account the bipartition $(m, M-m)$ of the modes,
\begin{equation}
|\psi\rangle=\sum_{\{k\},\{\alpha\}} C_{\{k\},\{\alpha\}}\ | k_1, \ldots, k_m\rangle\, 
| \alpha_{m+1}, \ldots, \alpha_M\rangle\ ,
\label{9}
\end{equation}
where the sums are over all integers $k_i$, $i=1,\ldots, m$ and $\alpha_j$, $j=m+1,\ldots,M$.
Inserting this decomposition in the separability condition (\ref{5-1})
and choosing 
$A_1=(a_1^\dagger)^{k'_1}\cdots (a_m^\dagger)^{k'_m} \cdot (a_1)^{k_1}\cdots (a_m)^{k_m}$
and $A_2=(a_{m+1}^\dagger)^{\alpha'_{m+1}} \cdots (a_{M}^\dagger)^{\alpha'_{M}}
 \cdot (a_{m+1})^{\alpha_{m+1}}\cdots (a_{M})^{\alpha_{M}}$,
one gets the following set of constraints
on the complex coefficients $C$ (the bar signifies complex conjugation):
\begin{equation}
\overline{C}_{\{k'\},\{\alpha'\}}\, C_{\{k\},\{\alpha\}}=
\sum_{\{\beta\}} \overline{C}_{\{k'\},\{\beta\}}\, C_{\{k\},\{\beta\}}\ 
\sum_{\{l\}} \overline{C}_{\{l\},\{\alpha'\}}\, C_{\{l\},\{\alpha\}}\ .
\label{10}
\end{equation}
Taking $k'=k$ and $\alpha'=\alpha$, this expression becomes
\begin{equation}
\big| C_{\{k\},\{\alpha\}} \big|^2= D_{\{k\}}\ D'_{\{\alpha\}}\ ,\qquad
D_{\{k\}}=\sum_{\{\beta\}} \big| C_{\{k\},\{\beta\}} \big|^2\geq0\ ,\quad
D'_{\{\alpha\}}=\sum_{\{l\}} \big| C_{\{l\},\{\alpha\}} \big|^2\geq0\ ,
\label{11}
\end{equation}
so that the modulus of the coefficient $C$ factorizes, and one can write:
\begin{equation}
C_{\{k\},\{\alpha\}}= \big(D_{\{k\}}\big)^{1/2}\ \big(D'_{\{\alpha\}}\big)^{1/2}\ 
e^{i\theta_{\{k\},\{\alpha\}}}\ ,\qquad
\sum_{\{k\}} D_{\{k\}}=1\ ,\quad
\sum_{\{\alpha\}} D'_{\{\alpha\}}=1\ .
\label{12}
\end{equation}
Insertion of this expression back into the condition (\ref{10}) produces
additional constraints for the phases $\theta$:
\begin{equation} 
e^{i( \theta_{\{k\},\{\alpha\}}-\theta_{\{k'\},\{\alpha'\}})}
=e^{i( \theta_{\{k\},\{\beta\}}-\theta_{\{k'\},\{\beta\}})}\
e^{i( \theta_{\{l\},\{\alpha\}}-\theta_{\{l\},\{\alpha'\}})}\ .
\label{13}
\end{equation}
From this expression one immediately sees that the phase difference
$(\theta_{\{k\},\{\beta\}}-\theta_{\{k'\},\{\beta\}})$ does not depend on 
the labels $\{\beta\}$ and vanishes for $k=k'$, 
while $(\theta_{\{l\},\{\alpha\}}-\theta_{\{l\},\{\alpha'\}})$
is independent from the labels $\{l\}$ and vanishes for $\alpha=\alpha'$.
With this information, one can rewrite the decomposition (\ref{9}) as
\begin{eqnarray}
\label{14}
|\psi\rangle&=&\sum_{\{k\}} \big(D_{\{k\}}\big)^{1/2}| k_1, \ldots, k_m\rangle\
\sum_{\{\alpha\}} \big(D'_{\{\alpha\}}\big)^{1/2}\, e^{i\theta_{\{k\},\{\alpha\}}}
| \alpha_{m+1}, \ldots, \alpha_M\rangle\\
&=&\sum_{\{k\}} \big(D_{\{k\}}\big)^{1/2}\, e^{i\theta_{\{k\},\{\alpha'\}}}| k_1, \ldots, k_m\rangle\
\sum_{\{\alpha\}} \big(D'_{\{\alpha\}}\big)^{1/2}\, e^{i(\theta_{\{k\},\{\alpha\}}-\theta_{\{k\},\{\alpha'\}})}
| \alpha_{m+1}, \ldots, \alpha_M\rangle\ ,
\nonumber
\end{eqnarray}
where $\{\alpha'\}$ is any given set of reference occupation numbers. However, as observed before,
the difference $(\theta_{\{k\},\{\alpha\}}-\theta_{\{k\},\{\alpha'\}})$ is independent
from $\{k\}$, so that the expression of $|\psi\rangle$ given above
is precisely of the form (\ref{7}).\hfill$\square$
\medskip

When the state of the bosonic many-body system is not pure, it can be described by a density
matrix $\rho$, that, using the Fock basis of (\ref{6}), can be written as:
\begin{equation}
\rho=\sum_{\{n\},\{n'\}}\rho_{\{n\},\{n'\}}\ |n_1, n_2,\ldots,n_M\rangle\langle n'_1, n'_2,\ldots,n'_M|\ ,
\quad \sum_{\{n\}}\rho_{\{n\},\{n\}}=1\ .
\label{15}
\end{equation}
The set of all states form a convex set, whose extremals are given by the pure ones.
{\sl Proposition 1} can then be used to characterize separable mixed states:

\medskip
{\bf Corollary 1.} {\sl A mixed state $\rho$ as in (\ref{15}) is
$({\cal A}_1,\, {\cal A}_2)$-separable if and only if it is the
convex combination of projectors on pure $({\cal A}_1,\, {\cal A}_2)$-separable
states; otherwise, the state $\rho$ is $({\cal A}_1,\, {\cal A}_2)$-entangled.}
\medskip 

Until now, we have considered many-body systems formed by an arbitrary number of elementary constituents;
we shall now restrict the discussion to systems with a given number $N$ of bosonic particles. 
In this case, the Hilbert space $\cal H$ becomes finite dimensional: it is still spanned
by the Fock states (\ref{6}); however, since
the total number operator $\sum_{i=1}^M a_i^\dagger a_i$
is a conserved quantity, the occupation numbers must satisfy the 
additional constraint $\sum_{i=1}^M n_i=N$, {\it i.e.} all states need
to contain exactly $N$ particles.

Note that in absence of this number ``superselection'' rule, the many-body Hilbert space $\cal H$
spanned by the Fock states (\ref{6}) possesses a {\sl mode} tensor product structure similar
to the {\sl particle} tensor product in (\ref{2}),
${\cal H}={\cal H}^{(1)}\otimes{\cal H}^{(2)}\otimes\ldots\otimes{\cal H}^{(M)}$,
where ${\cal H}^{(i)}$ now represents the $i$-th mode Fock space spanned by the
states $|n_i\rangle \sim (a_i^\dagger)^{n_i} |0\rangle$, $n_i\in \mathbb{N}$.
This structure is clearly lost when the number of the system constituents is fixed,
since now all states must contain the same number of bosons.

The conservation of the total particle number has consequences also on the form taken by pure separable
states $|\psi\rangle$; they still have to be in product form as in (\ref{7}), but now ${\cal P}$ and ${\cal Q}$
need to be homogeneous polynomials of complementary degree:
if ${\cal P}$ is of degree $k$, than ${\cal Q}$ must be of degree $N-k$. In other terms,
given the mode partition $(m, M-m)$ leading to the bipartition $({\cal A}_1,\, {\cal A}_2)$ of
the whole algebra ${\cal B}({\cal H})$,
if there are $k$ particles filling the first $m$ modes, then there must be $N-k$ in the
remaining $M-m$ modes; only in this way the state $|\psi\rangle$ will contain the prescribed $N$ bosons.

Examples of $N$ bosons pure separable states are the Fock states; indeed,
using the notation and specifications introduced before, they can be recast in the form (\ref{7}):
\begin{equation}
| k_1, \ldots, k_m\rangle\, | \alpha_{m+1}, \ldots, \alpha_M\rangle\ ,\quad
\sum_{i=1}^m k_i =k\ ,\ \sum_{j=m+1}^M \alpha_j=N-k\ ,\ \ 0\leq k \leq N\ ;
\label{16}
\end{equation}
by varying the integers $k_i$, $\alpha_j$ and $k$, these states generate the whole Hilbert space $\cal H$.
This basis states can be relabeled in a different, more convenient way as:
\begin{equation}
| k, \sigma; N-k, \sigma'\rangle\ ,\quad \sigma=1,2, \ldots, {k+m-1\choose k}\ ,\
\sigma'=1, 2,\ldots, {N-k+M-m-1\choose N-k}\ ;
\label{17}
\end{equation}
as before the integer $k$ represents the number of particles found in the first $m$ modes,
while $\sigma$ counts the different ways in which those particles can fill those modes;
similarly, $\sigma'$ labels the ways in which the remaining $N-k$ particles
can occupy the other $M-m$ modes. Clearly, due to the orthonormality of the states
in (\ref{16}), in this new labelling one has:
$\langle k, \sigma; N-k, \sigma'| l, \tau; N-l, \tau'\rangle=
\delta_{kl}\,\delta_{\sigma\tau}\,\delta_{\sigma'\tau'}$.
Using this notation, a generic mixed state $\rho$
can then be written as:
\begin{equation}
\rho=\sum_{k,l=0}^N\ \sum_{\sigma,\sigma',\tau,\tau'}\
\rho_{k \sigma\sigma', l\tau\tau'}\ | k, \sigma; N-k, \sigma'\rangle \langle l, \tau; N-l, \tau' |\ ,
\quad \sum_{k=0}^N\ \sum_{\sigma,\sigma'}\
\rho_{k \sigma\sigma', k\sigma\sigma'}=1\ .
\label{18}
\end{equation}
Different orthonormal basis of separable pure states can be chosen for $\cal H$
and will be used in the following; they are all characterized by the integer $k$ that
fixes the number of particles filling the first $m$ modes, plus some additional
indices able to distinguish orthogonal states with the same number of particles
in the two partitions.

\section{The partial transposition criterion}

In the previous Section, we have seen that, given a bipartition $({\cal A}_1,\, {\cal A}_2)$
of the $N$-boson algebra, obtained through the $(m, M-m)$-partition of the $M$ modes, 
separable pure states containing $k$ particles in the first modes (and
$N-k$ in second ones) are in product form as given in (\ref{7}), or equivalently, when expressed
in the basis of the Fock states (\ref{17}):
\begin{equation}
|\psi\rangle=\sum_{\sigma,\sigma'} 
C_\sigma^{(k)}\, {C'}_{\sigma'}^{(k)}\ | k, \sigma; N-k, \sigma'\rangle\ ,
\label{19}
\end{equation}
while separable mixed states are convex combinations of their projectors. However, to determine
whether a given density matrix can be written in separable form is in general a hard task.
In the case of systems of distinguishable particles, easily implementable separability tests have been
constructed. In the case of bipartite systems, one of the most useful criteria involves the operation
of partial transposition \cite{Peres,Horodecki2}: 
a state $\rho$ for which the partially transposed density matrix $\tilde\rho$ is
no longer positive is surely entangled. This lack of positivity can be witnessed by the
so-called negativity \cite{Zyczkowski,Vidal,Horodecki}:
\begin{equation}
{\cal N}(\rho)=\, {1\over2}\Big(\!\parallel\tilde\rho\parallel_1 - {\rm Tr}[\rho]\Big)\ ,\qquad 
\parallel\tilde\rho\parallel_1={\rm Tr}\Big[\sqrt{\tilde\rho^\dagger \tilde\rho}\Big]\ .
\label{20}
\end{equation}
Indeed, if $\tilde\rho$ is not positive, 
then $\parallel\tilde\rho\parallel_1>{\rm Tr}[\rho]={\rm Tr}[\tilde\rho]=1$ 
and as consequence \hbox{${\cal N}(\rho)>0$}.
Unfortunately, this criterion is not exhaustive, since there are entangled states that remain
positive under partial transposition.
The negativity is nevertheless an exhaustive entanglement witness for systems composed
by two qubits or one qubit and one qutrit \cite{Peres,Horodecki2},
or, in continuous variable systems, for two-mode Gaussian states \cite{Simon}.

We shall now examine to what extent these results, valid in the case of bipartite systems of
distinguishable particles, can be extended to systems composed by $N$ bosons;
we will see that the identity of the particles induces novel features in the theory of
many-body entanglement.

We shall work within the generic $({\cal A}_1,\, {\cal A}_2)$-bipartition introduced above,
induced by the $(m, M-m)$-splitting of the $M$ modes. First notice that the operation
of partial transposition ({\it e.g.} with respect to the first subalgebra) maps separable states
into separable states. This is immediate for pure states as in (\ref{19}), 
since in this case partial transposition
amounts to the complex conjugation of the first coefficient,
$C_\sigma^{(k)}\mapsto \overline{C}_\sigma^{(k)}$; being convex combinations of projection on
pure separable states, also separable mixed states are similarly mapped into perfectly admissible
density matrices. As a consequence, also in the case of bosonic systems, negativity of a partially
transposed density matrix would definitely signal $({\cal A}_1,\, {\cal A}_2)$-entanglement.

For a generic density matrix $\rho$ as in (\ref{18}), the computation of the negativity
${\cal N}(\rho)$ involves the evaluation of the trace norm $\parallel\tilde\rho\parallel_1$
of the partially transposed matrix:
\begin{equation}
\tilde\rho=\sum_{k,l=0}^N\ \sum_{\sigma,\sigma',\tau,\tau'}\
\rho_{k \sigma\sigma', l\tau\tau'}\ | l, \tau; N-k, \sigma'\rangle \langle k, \sigma; N-l, \tau' |\ ,
\label{21}
\end{equation}
usually a rather difficult task (however, see Section 5 below). 
Nevertheless, there are relevant, quite general instances in which
the evaluation of ${\cal N}(\rho)$ can be performed analytically. 

As a prelude to more complex situations, we shall first discuss the case of
a system of $N$ bosons that can occupy just two modes, {\it i.e.} $M=2$,
each of the two algebras ${\cal A}_1$ and ${\cal A}_2$ being generated by the creation and annihilation
operators in just one of the two modes;
this example is quite important in applications since it is the paradigm for the description of
ultracold bosonic gases confined in double well potentials \cite{Milburn,Jaksch}.
One finds that in this case the negativity is an exhaustive entanglement witness, as
stated by the following:

\medskip
{\bf Lemma 1.} {\sl A two-mode state of an $N$-boson system is entangled with respect to the 
bipartition into single modes if and only if its partial transposition is negative.}
\medskip 

{\sl Proof.} A generic state for a two mode $N$-boson system can be written
as in (\ref{18}), although in a simplified form:
\begin{equation}
\rho=\sum_{k,l=0}^N\ 
\rho_{k, l}\ | k; N-k\rangle \langle l ; N-l |\ , \qquad \sum_{k,l=0}^N\ \rho_{k, k}=1\ ,
\label{22}
\end{equation}
since in this case a single integer is sufficient to label the $N+1$-dimensional basis of Fock states.
The partial transposition operator (with respect to the first mode) maps this state to an
operator $\tilde\rho$ on a larger Hilbert space than that with a fixed $N$,
\begin{equation}
\tilde\rho=\sum_{k,l=0}^N\ 
\rho_{k, l}\ | l; N-k\rangle \langle k ; N-l |\ ,
\label{23}
\end{equation}
such that the combination
\begin{equation}
\tilde\rho^\dagger \tilde\rho=\sum_{k,l=0}^N\ 
|\rho_{k, l}|^2\ | k; N-l\rangle \langle k ; N-l |
\label{24}
\end{equation}
is diagonal. Its trace norm can then be readily computed, as the negativity in
(\ref{20}):
\begin{equation}
{\cal N}(\rho)={1\over 2}\sum_{k\neq l=0}^N\ |\rho_{k, l}|\ .
\label{25}
\end{equation}
This expression vanishes if and only if the state is diagonal in the Fock basis,
but this is equivalent to separability, since in this case $\rho$ results a convex
combinations of projectors over Fock states.\hfill$\square$

These conclusions can be extended to the case of bosonic systems in $M$ modes
by choosing a $({\cal A}_1,\, {\cal A}_2)$-bipartition for which the algebra
${\cal A}_1$ is generated by creation and annihilation operators of one single mode,
while the remaining $M-1$ modes generates ${\cal A}_2$. Also in this case
the negativity results an exhaustive entanglement criterion as stated in:

\medskip
{\bf Proposition 2.} {\sl A $M$-mode state of an $N$-boson system is entangled with respect to the 
the bipartition into one mode and the rest if and only if its partial 
transposition is negative.}
\medskip 

{\sl Proof.} In this case, a generic state of the system can be written as
in (\ref{18}), dropping however the first greek labels in the Fock states:
\begin{equation}
\rho=\sum_{k,l=0}^N\ \sum_{\sigma,\tau}\
\rho_{k \sigma, l\tau}\ | k; N-k, \sigma\rangle \langle l; N-l, \tau |\ ,
\quad \sum_{k=0}^N\ \sum_{\sigma}\
\rho_{k \sigma, k\sigma}=1\ .
\label{26}
\end{equation}
The operation of partial transposition with respect to the first algebra can be readily implemented
\begin{equation}
\tilde\rho=\sum_{k,l=0}^N\ \sum_{\sigma,\tau}\
\rho_{k \sigma, l\tau}\ | l; N-k, \sigma\rangle \langle k; N-l, \tau |\ ,
\label{27}
\end{equation}
so that
\begin{equation}
\tilde\rho^\dagger \tilde\rho=\sum_{k,l=0}^N\ \sum_{\sigma,\tau}\
\big|\rho_{k \sigma, l\tau}\big|^2\ 
| \phi_{kl}^\sigma\rangle \langle \phi_{kl}^\sigma|\ ,
\label{28}
\end{equation}
where
\begin{equation}
|\phi_{kl}^\sigma\rangle=
\bigg(\sum_{\tau}\big|\rho_{k \sigma, l\tau}\big|^2\bigg)^{-1/2}\ 
\sum_{\sigma'}\
\overline{\rho}_{k \sigma, l\sigma'}\ | k; N-l, \sigma'\rangle\ .
\label{29}
\end{equation}
By separating diagonal and off-diagonal terms in $k$ and $l$, the positive matrix in (\ref{28}) can
be further decomposed as
\begin{eqnarray}
\label{30}
&&\tilde\rho^\dagger \tilde\rho=D+R\ ,\\
\label{31}
&&D=\sum_{k=0}^N\ \sum_{\sigma}\ 
\bigg(\sum_{\tau}\big|\rho_{k \sigma, k\tau}\big|^2\bigg)\
| \phi_{kk}^\sigma\rangle \langle \phi_{kk}^\sigma|\ ,\\
\label{32}
&&R=\sum_{k\neq l=0}^N\ \sum_{\sigma}\
\bigg(\sum_{\tau}\big|\rho_{k \sigma, l\tau}\big|^2\bigg)\ 
| \phi_{kl}^\sigma\rangle \langle \phi_{kl}^\sigma|\ .
\end{eqnarray}
Thanks to the orthogonality of the vectors $| \phi_{kl}^\sigma\rangle$ for different indices
$k$ or $l$, one finds that the products $DR$ and $RD$ vanish, and thus
one has:
\begin{equation}
\sqrt{\tilde\rho^\dagger \tilde\rho}=\sqrt{D} + \sqrt{R}\ .
\label{33}
\end{equation}
In order to explicitly compute $\sqrt{D}$, we make a change of basis in the second $M-1$ partition
passing from the Fock states to another set of separable states adapted to $\rho$,
such that its components along this new basis satisfy: 
$\rho_{k \sigma, k\tau}=\rho_{k \sigma, k\sigma}\ \delta_{\sigma\tau}$; notice that this is always possible
through suitable local, unitary transformations diagonalizing for each $k$ the matrices
${\cal M}_{\sigma\tau}^{(k)}\equiv\big[\rho_{k \sigma, k\tau}\big]$. In this new basis,
the vectors $|\phi_{kl}^\sigma\rangle$ become orthonormal in all three indices,
and as a consequence, recalling (\ref{31}),
\begin{equation}
\sqrt{D}=
\sum_{k=0}^N\ \sum_{\sigma}\ \big|\rho_{k \sigma, k\sigma}\big|\
| \phi_{kk}^\sigma\rangle \langle \phi_{kk}^\sigma|\ ;
\label{34}
\end{equation}
further, since the diagonal elements $\rho_{k \sigma, k\sigma}$ are non-negative, 
one has: ${\rm Tr}\big[\sqrt{D}\big]=$\break
$\sum_{k=0}^N\ \sum_{\sigma}\ \rho_{k \sigma, k\sigma}=1$, being $\rho$ normalized;
it follows that the negativity (\ref{20}) of the general state (\ref{26}) is simply given by
\begin{equation}
{\cal N}(\rho)={1\over 2}{\rm Tr}\Big[\sqrt{R}\Big]\ .
\label{35}
\end{equation}
Recalling the explicit expression of $R$ in (\ref{32}), 
one immediately sees that the negativity vanishes
if and only if the elements $\rho_{k \sigma, l\tau}$ for $k\neq l$ are all zero.
But this means that $\rho$, expressed in the new adapted basis, is the convex sum
of projectors on separable pure states.\hfill$\square$

In the above proof we have used the partial transposition operation with
respect to the first party. The same conclusion nevertheless holds by choosing
the partial transposition operation with respect to the second party, since
it can be obtained as the combination of the partial transposition with respect to the first party
followed by the total transposition, and this second operation does not
change the negativity.

The previous proof suggests the existence of a larger class of bipartite states 
for which the negativity of the partial transposed
states uniquely signals the presence of entanglement.

\medskip
{\bf Corollary 2.} {\sl Let us consider the class of $(m,M-m)$
mode bipartite states (\ref{18}) whose principal minors
$\rho_{k \sigma\sigma', k\tau\tau'}$ for each $k$ turn out to be diagonal
in a basis of separable pure vectors. Then, these states result entangled with respect
to the fixed bipartition if and only if their partial transposition is negative.}
\medskip 

{\sl Proof.} This statement can be proved through a straightforward generalization 
of the steps followed in proving {\sl Proposition 2}. In the chosen basis, one
can decompose a generic state $\rho$ as in (\ref{18}). Then, by direct computation,
one arrives at a decomposition $\tilde\rho^\dagger \tilde\rho=D+R$ similar to the one
in (\ref{30})-(\ref{32}); the class of considered states are such that
$\rho_{k \sigma\sigma', k\tau\tau'}\propto \delta_{\sigma\tau}\, \delta_{\sigma' \tau'}$,
and this allows to conclude that ${\rm Tr}\big[\sqrt{D}\big]=1$. Then, as before, one finds that
the negativity ${\cal N}(\rho)={\rm Tr}\big[\sqrt{R}\big]/2$ is zero if and only if
the off-diagonal elements $\rho_{k \sigma\sigma', l\tau\tau'}$, for $k\neq l$, vanish, {\it i.e.}
if and only if the state $\rho$ is separable.\hfill$\square$

\section{States remaining positive under partial transposition}

Although the results proven so far cover a large set of bosonic states, 
the negativity of a partially transposed state
is not in general a necessary condition for entanglement: there are
states that remain positive under partial transposition (they are usually called PPT states), 
but nevertheless turn out to be entangled. As a first step towards their classification, it
is important to give an explicit characterization of $N$-boson states that are PPT;
this can be obtained by means of techniques similar to those used above.

\medskip
{\bf Proposition 3.} {\sl A general $(m, M-m)$-mode bipartite $N$-boson state (\ref{18})
is positive under partial transposition (PPT) if and only if it is
``block diagonal'', namely of the form
\begin{equation}
\rho=\sum_{k=0}^N \rho_k\ ,
\label{36}
\end{equation}
with
\begin{equation}
\rho_k=\sum_{\sigma,\sigma',\tau,\tau'}\
\rho_{k \sigma\sigma', k\tau\tau'}\ | k, \sigma; N-k, \sigma'\rangle \langle k, \tau; N-k, \tau' |\ ,
\label{37}
\end{equation}
and its principal minors $\rho_k$ are PPT.}
\medskip 

{\sl Proof.} Given the decomposition (\ref{18}) of a generic state $\rho$ in the Fock basis
adapted to the chosen $(m, M-m)$-bipartition and its partial transposition (\ref{21}),
one sees that the positive operator $\tilde\rho^\dagger\tilde\rho$ can be cast in the form:
\begin{equation}
\tilde\rho^\dagger \tilde\rho=\sum_{k,l=0}^N\ \sum_{\sigma,\tau}\
\bigg(\sum_{\sigma'\tau'}\big|\rho_{k \sigma'\tau,l \sigma\tau'}\big|^2\bigg)\ 
| \Phi_{kl}^{\sigma\tau}\rangle \langle \Phi_{kl}^{\sigma\tau}|\ ,
\label{38}
\end{equation}
where the vectors
\begin{equation}
|\Phi_{kl}^{\sigma\tau}\rangle=
\bigg(\sum_{\sigma'\tau'}\big|\rho_{k \sigma'\tau,l \sigma\tau'}\big|^2\bigg)^{-1/2}\ 
\sum_{\sigma'\tau'}\
\overline{\rho}_{k \sigma'\sigma, l\tau\tau'}\ | k, \sigma'; N-l, \tau'\rangle\ ,
\label{39}
\end{equation}
are orthogonal with respect to the latin indices, but in general not in the greek ones. 
Nevertheless, this is sufficient for proving that the following decomposition:
\begin{equation}
\sqrt{\tilde\rho^\dagger \tilde\rho}=\sum_{k=0}^N\sqrt{\cal D}_k + \sqrt{\cal R}\ ,
\label{40}
\end{equation}
similar to the one encountered before in a simpler setting, holds; explicitly, one has:
\begin{eqnarray}
\label{41}
&&{\cal D}_k= \sum_{\sigma,\tau}\ 
\bigg(\sum_{\sigma'\tau'}\big|\rho_{k \sigma'\tau, k\sigma\tau'}\big|^2\bigg)\ 
| \Phi_{kk}^{\sigma\tau}\rangle \langle \Phi_{kk}^{\sigma\tau}|\ ,\\
\label{42}
&&{\cal R}=\sum_{k\neq l=0}^N\ \sum_{\sigma,\tau}\
\bigg(\sum_{\sigma'\tau'}\big|\rho_{k \sigma'\tau, l\sigma\tau'}\big|^2\bigg)\ 
| \Phi_{kl}^{\sigma\tau}\rangle \langle \Phi_{kl}^{\sigma\tau}|\ ,
\end{eqnarray}
so that one immediately checks that ${\cal D}_k\, {\cal D}_{k'}=\,0$ for $k\neq k'$ and also
${\cal D}_k\, {\cal R}=0={\cal R}\,{\cal D}_k$.
Further, by direct computation, one finds that ${\cal D}_k=(\tilde\rho_k)^\dagger\tilde\rho_k$,
where $\rho_k$ is the following operator built with the ``block-diagonal'' 
components of $\rho$ ({\it cf.} (\ref{37})):
\begin{equation}
\rho_k=\sum_{\sigma,\sigma',\tau,\tau'}\
\rho_{k \sigma\sigma', k\tau\tau'}\ | k, \sigma; N-k, \sigma'\rangle \langle k, \tau; N-k, \tau' |\ ,
\qquad \sum_{k=0}^N{\rm Tr}[\rho_k]=1\ ,
\nonumber
\end{equation}
the second expression coming from the normalization of $\rho$.
As a consequence, the negativity of the generic $N$-boson state $\rho$ can be written as:
\begin{equation}
{\cal N}(\rho)=\sum_{k=0}^N {\cal N}(\rho_k) + {1\over 2}{\rm Tr}\Big[\sqrt{\cal R}\Big]\ ,
\label{43}
\end{equation}
in terms of the negativities ${\cal N}(\rho_k)$ of the operators $\rho_k$.
Since $\cal R$ as given in (\ref{42}) is the (direct) sum of positive operators, ${\cal N}(\rho)$
in (\ref{43}) vanishes if and only if each term does, namely $\rho$ is precisely of the form (\ref{36}),
(\ref{37}), and further ${\cal N}(\rho_k)=\,0$.\hfill$\square$

This complete characterization of the $N$-boson PPT states clearly simplify the search for
bound entangled states; indeed, the identification of entangled bosonic many-body states that
remain positive under partial transposition, in general a very difficult task, is reduced
by {\sl Proposition 3} to the more manageable problem of identifying quantum correlations
in their diagonal blocks (\ref{37}), for which standard techniques can be used \cite{Horodecki}.

\section{Applications and examples}

The result in (\ref{43}) gives the expression for the negativity of a generic 
$N$-boson state $\rho$, once written in terms of the Fock basis
adapted to the chosen bipartition as in (\ref{18}); from its explicit form 
one can appreciate that the statements 
of {\sl Proposition 2} and {\sl Corollary 2} proven before require much more stringent conditions 
than the ones assumed in the previous {\sl Proposition 3}, specifically, 
that the negativity of all diagonal blocks of $\rho$ be identically zero, ${\cal N}(\rho_k)=\,0$. 
In this respect, the conclusions of {\sl Proposition 3} are therefore more general,
but weaker, than the ones dealing with the partial transposition criterion in Section 4.

Yet, the general expression (\ref{43}) of the negativity
gives some additional information on entangled $N$-boson states in general;
they can be summarized by the following:

\medskip
{\bf Corollary 3.} {\sl If a generic $(m, M-m)$-mode bipartite state (\ref{18}) can 
not be cast in block diagonal form as in (\ref{36}), (\ref{37}), 
or if it can, at least one of its diagonal blocks is not PPT, then the state is non-separable
and its entanglement can be detected by the partial transposition criterion.}
\medskip

In other terms, a given many-body systems is very likely to be found in a state 
having non-vanishing off-diagonal elements
and therefore in an entangled state, since in this case ${\rm Tr}\Big[\sqrt{\cal R}\Big]\neq0$,
giving rise to a nonvanishing negativity, ${\cal N}(\rho)>0$.
Indeed, this is the case in many physical situations involving quantum phase transitions
in condensed and superconducting systems \cite{Sachdev},
where the so-called ``off-diagonal long range order'' 
states play a fundamental role ({\it e.g.} see \cite{Yang}-\cite{Giorda}).

Nevertheless, also block-diagonal states of the form (\ref{36}), (\ref{37}) can be encountered
in the study of many-body systems, due to the presence of noise. Indeed, actual many-body systems, 
albeit carefully screened, can not be considered totally isolated from the external
environment; its presence unavoidably produces noise and dissipation.%
\footnote{For an introduction to the theory of open quantum systems,
see \cite{Alicki}-\cite{Benatti3}.}
The most basic (and ubiquitous) effect induced by the environment is the so-called ``dephasing noise'',
leading to decoherence, {\it i.e.} to the suppression of the off-diagonal elements 
in the system density matrix.

More specifically, in our $N$-boson setting, let us consider a general $(m, M-m)$-mode bipartition
of our system, initially prepared in a generic state $\rho$: it can be decomposed as in (\ref{18}) 
in terms of the Fock states $| k, \sigma; N-k, \sigma'\rangle$ containing $k$ bosons in the first partition
and $N-k$ in the remaining one. Using the mode creation and annihilation operators introduced
at the beginning of Section 3, one can now consider the following collective observable
\begin{equation}
V=\sum_{i=1}^m a^\dagger_i a_i - \sum_{\alpha=m+1}^M a^\dagger_\alpha a_\alpha\ ,
\label{44}
\end{equation}
that counts the relative number of bosons in the two partitions; it clearly commutes
with the total number operator $\sum_{i=1}^M a^\dagger_i a_i \equiv N$. 
The effects of dephasing noise can then be described by the following 
time-evolution master equation \cite{Alicki}-\cite{Benatti3}:
\begin{equation}
{\partial\rho(t)\over\partial t}={\gamma\over2}\bigg(V\rho(t) V-{1\over2}\Big\{V^2,\rho(t)\Big\}\bigg)\ ,
\label{45}
\end{equation}
where the positive coupling constant $\gamma$ measures the strength of the noise.
Since the Fock states are eigenstates of $V$, 
$V\,| k, \sigma; N-k, \sigma'\rangle=(2k-N)\, | k, \sigma; N-k, \sigma'\rangle$,
one easily finds that the matrix elements
$\rho_{k\sigma\sigma',l,\tau\tau'}(t)=\langle k, \sigma; N-k, \sigma'|\rho(t)| l, \tau; N-l, \tau'\rangle$
obey:
\begin{equation}
{\partial\over\partial t}\,\rho_{k\sigma\sigma',l,\tau\tau'}(t)=
-\gamma\, (k-l)^2\,\rho_{k\sigma\sigma',l,\tau\tau'}(t)\ .
\label{46}
\end{equation}
It thus follows that this kind of noisy, irreversible dynamics indeed leads to the suppression of the
entries that are off-diagonal in the partitions occupation number:
\begin{equation}
\rho(t)=\sum_{k,l=0}^N\ \sum_{\sigma,\sigma',\tau,\tau'}\ e^{-\gamma t\,(k-l)^2}\ 
\rho_{k \sigma\sigma', l\tau\tau'}\ | k, \sigma; N-k, \sigma'\rangle \langle l, \tau; N-l, \tau' |\ .
\label{47}
\end{equation}
As a consequence, in the long time regime, $\rho(t)$ becomes diagonal with respect
to this occupation number, the faster, the larger the coupling $\gamma$ is; at the end,
$\rho$ takes precisely 
the block-diagonal form given in (\ref{36}), (\ref{37}).

An experimentally relevant instance in which such diagonalization may rapidly occur
is that of a system of $N$ bosons with an internal (two-dimensional) hyperfine degree of freedom, confined
in a double well potential:%
\footnote{For actual realizations, see \cite{Gross,Riedel}.}
any boson can be found in one of the two wells
with one of the two possible internal ``polarizations'', so that in this case $M=4$.
A natural bipartition is the spatial one defined by the two wells (giving $m=2$).
The associated Fock basis states $| k, \sigma; N-k, \sigma'\rangle$
are labelled by the integer $k$ giving the number of bosons in the first well ($N-k$ are in the remaining one)
and by the additional integers $\sigma$, $\sigma'$, {\it e.g.} labelling the number of bosons
in the hyperfine ground state in the first, respectively second well; recalling (\ref{17}), 
at fixed $k$, these states generate an Hilbert space isomorphic to
$\mathbb{C}^{k+1}\otimes\mathbb{C}^{N-k+1}$. 

The dephasing noise
acts now on bosons in different wells, destroying the coherence between the ``spatial'' modes, leaving
instead untouched the internal, hyperfine ones. The resulting density matrix is block-diagonal,
$\rho=\sum_k \rho_k$, where each block $\rho_k$ is a (non normalized) density matrix
on $\mathbb{C}^{k+1}\otimes\mathbb{C}^{N-k+1}$. Clearly, as stated by {\sl Corollary 3}, a non vanishing
negativity of one of the $\rho_k$'s would signal entanglement for the $N$-body state described
by the total density matrix $\rho$. Furthermore, note that, thanks to {\sl Proposition 3},  
PPT states $\rho$ that are entangled are also possible, provided $N\geq4$:
indeed, recalling the results of \cite{Peres, Horodecki2},
only in this case the matrix $\rho$ can accommodate diagonal blocks $\rho_k$ large enough
to allow PPT entanglement.

As an explicit example, consider the system composed by four bosons ($N=4$), that can occupy
four different modes ($M=4$), and choose $m=2$, so that each partition is made of two modes.
The block diagonal density matrix $\rho$ in (\ref{36}) contains now five terms.
Each diagonal block $\rho_k$ is fully described by the corresponding coefficient
matrix $\rho_{k\sigma\sigma',k\tau\tau'}$: for $k=0,4$ these matrices are $5\times 5$,
$8\times8$ when $k=1,3$, while in the case $k=2$ it turns out to be $9\times9$.
Let us focus on this last $\rho_{k=2}$ block:
\begin{equation}
\rho_{k=2}=\sum_{\sigma,\sigma',\tau,\tau'}\
\rho_{ \sigma\sigma', \tau\tau'}\ |\sigma;\sigma'\rangle \langle \tau;\tau' |\ ,
\end{equation}
where on the r.h.s. we have suppressed for simplicity the labels $k=N-k=2$. Since now
$\sigma,\sigma',\tau,\tau'=1,2,3$, the Hilbert subspace spanned by the set of six
Fock vectors $\{|\sigma;\sigma'\rangle,\ \sigma,\sigma'=1,2,3\}$
is isomorphic to $\mathbb{C}^{3}\otimes\mathbb{C}^{3}$, {\it i.e.} to a two
qutrit space. For such a system, a bound entangled state has been
presented in \cite{Horodecki3} and its $9\times9$ density matrix explicitly
written in the $\mathbb{C}^{3}\otimes\mathbb{C}^{3}$ standard basis.
By identifying the six Fock vectors $\{|\sigma;\sigma'\rangle\}$ with such a basis,
one can choose for the coefficient matrix $\rho_{ \sigma\sigma', \tau\tau'}$
precisely that PPT entangled density matrix. 
Assuming that the remaining blocks $\rho_k$, $k=1,3$, are separable 
(the other two blocks $\rho_k$, $k=0,4$ are always separable),
the density operator $\rho=\sum_{k=0}^4 \rho_k$ results
a bound entangled state for the system.

As a final remark, notice that all above results can be generalized to the case
of systems where the total number of particles is not fixed,
but commutes with all physical observables ({\it i.e.} we are in
presence of a superselection rule \cite{Bartlett}).
In such a situation, general density matrices are incoherent mixtures 
of states of the form (\ref{18}) with fixed number $N$ of
particles:
\begin{equation}
\rho=\sum_N \lambda_N \rho_N \,\qquad \lambda_N\geq 0\ ,\qquad \sum_N \lambda_N=1\ .
\end{equation}
In this case, a suitable entanglement measure is the weighted sum of the negativities
of each component $\rho_N$, {\it i.e.} $\sum_N \lambda_N\,  {\cal N}(\rho_N)$.
All the arguments discussed in the previous sections apply to the single $\rho_N$,
and therefore the statements of {\sl Lemma 1} and {\sl Propositions 2, 3} also hold
in this more general setting.%
\footnote{Being in presence of the number superselection rule, admissible pure states of the system
need to be eigenvectors of the total number operator, while mixed states must be decomposed
as their convex combinations. In general, the partial transposition operation mixes sectors
with different $N$, giving rise to positive partially transposed matrices
for states $\rho$ not satisfying the separability criterion (\ref{3});
indeed, this happens for states $\rho$ whose separable convex decompositions
always involve superpositions of different total number eigenspaces. Therefore, in presence of
incoherent fluctuations of the number operator, the negativity as defined in (\ref{20}) 
does not appear to be a useful entanglement measure.}
At the same time, the effects of the dephasing noise
given by the dynamics (\ref{45}) are still the asymptotic block diagonalization of each $\rho_N$.

Much more difficult is the treatment of the case of
many-body states with fluctuating number of particles,
{\it i.e.} of density matrices that are coherent mixtures 
of states with different $N$.
At least when the number fluctuation is small, one can nevertheless
advance some general considerations, based on the property of continuity
satisfied by the negativity in (\ref{20}).
If one starts with a $N$-boson state with strictly positive negativity 
({\it i.e.} an entangled state), by the permanence of sign, 
perturbations by states with small number fluctuation 
should not be able to turn it into a separable state.
On the contrary, any similar small perturbation of
a separable $N$-boson state ({\it i.e.} with vanishing negativity)
will in general turn it into an entangled state,
although with infinitesimal negativity;
this is so because after perturbation the state will no longer be
in block diagonal form.

\section{Outlook}

In the case of bosonic many-body systems composed by a fixed number of identical particles
the usual notions of separability and entanglement are inapplicable since 
the Hilbert space particle tensor product structure on which they are based is no longer available.
A generalized definition of separability valid in all physical situations
can however be formulated: its based on commuting algebras of observables instead of 
the particle tensor decomposition of states. This new framework makes it apparent that
in systems of identical particles, the notions of separability and entanglement
are meaningful only in reference to a specific choice of commuting sets of observables.

Using this generalized definition, we have studied bipartite entanglement in systems composed
by $N$ identical bosons that can occupy $M$ different modes.
More specifically, we have analyzed to what extent the partial transposition operation
can be used to detect quantum correlations in such systems and found that
in general it is a much more exhaustive criterion for detecting bipartite entanglement than in the case 
of systems of distinguishable particles.
Indeed, by giving explicit expressions to the negativity, the quantity that measures
the lack of positivity of the partially transposed density matrix, we have shown
that the operation of partial transposition gives a necessary and sufficient test of entanglement detection
for very general classes of bosonic states.

Even more relevant is the fact that such classes of states find direct applications
in many physical models. 
We have already remarked that the two-mode case ($M=2$)
represents the paradigm for discussing quantum effects in the behaviour of ultracold gases of bosonic atoms
confined in double-well traps; these systems have recently attracted a lot
of attention in view of their metrological applications \cite{Helstrom}-\cite{Paris}, and in particular 
in relation to the possibility of constructing interferometric devices able to overcome
the so-called shot-noise limit in parameter estimation \cite{Caves}-\cite{Oberthaler}.

Similarly, systems with several bosonic modes are used to model ultracold gases in optical lattices
and other many-body condensed systems \cite{Lewenstein}-\cite{Amico}, 
\cite{Stringari}-\cite{Yukalov},
in particular in reference to quantum phase transitions \cite{Sachdev}.
For such systems, the classes of entangled states discussed in {\sl Corollary 2, 3}
find application as ``off-diagonal long range order'' states
\cite{Yang}-\cite{Giorda}.

Even the block diagonal states analyzed in {\sl Proposition 3} are of relevance
in actual realizations. Indeed, as discussed in the last Section, they are the
natural outcome of dephasing noise; this makes quite realistic the possibility of
producing PPT entangled states in actual experimental apparata, {\it e.g.} those
described in \cite{Gross,Riedel}, through the action
of a suitably engineered environment.

\end{document}